\documentclass[fleqn,twoside]{article}

\usepackage{axodraw}
\usepackage{color}

\usepackage{espcrc2}
\usepackage{graphicx}

\readRCS
$Id: espcrc2.tex,v 1.2 2004/02/24 11:22:11 spepping Exp $
\usepackage{graphicx}
\usepackage[figuresright]{rotating}

\newcommand{\AmS}{{\protect\the\textfont2
  A\kern-.1667em\lower.5ex\hbox{M}\kern-.125emS}}

\title{
\vspace*{-35mm}
\rightline{
{\normalsize{DESY 04-135}}}
\vspace*{-2mm}
\rightline{
{\normalsize{NIKHEF 2004-008}}}
\vspace*{-2mm}
\rightline{\normalsize{August 2004}}
\vspace*{+20mm}
The QCD Splitting Functions at Three Loops: Methods and Results{\thanks{Presented 
by S.M. and J.A.M.V. at {\it{Loops and Legs in Quantum Field Theory}},
25th - 30th April 2004, Zinnowitz (Germany).}}}

\author{S. Moch\address{Deutsches Elektronensynchrotron DESY \\
Platanenallee 6, D--15735 Zeuthen, Germany},
J.A.M. Vermaseren\address[NIKHEF]{NIKHEF Theory Group \\
Kruislaan 409, 1098 SJ Amsterdam, The Netherlands}
 and A. Vogt\addressmark[NIKHEF]}
       
\newcommand{\beq}{\begin{equation}}
\newcommand{\eeq}{\end{equation}}
\newcommand{\bea}{\begin{eqnarray}}
\newcommand{\eea}{\end{eqnarray}}
\newcommand{\nn}{\nonumber}
\newcommand{\MSb}{$\overline{\mbox{MS}}$}
\newcommand{\as}{\alpha_{\rm s}}
\newcommand{\ra}{\rightarrow}

\newcommand{\gsim}{\raisebox{-0.07cm}{$\:\:\stackrel{>}{{\scriptstyle
 \sim}}\:\: $} }
\newcommand{\lsim}{\raisebox{-0.07cm}{$\:\:\stackrel{<}{{\scriptstyle
 \sim}}\:\: $} }

\begin{document}

\begin{abstract}
We have computed the complete next-to-next-to-leading order (NNLO) contributions 
to the splitting functions governing the evolution of unpolarized 
parton densities in perturbative QCD.
Our results agree with all partial results available in the literature.
We illustrate the methods used for this calculation with some examples and 
display selected results to show the size of the NNLO corrections and 
their effect on the evolution.
\vspace{1pc}
\vspace*{-4mm}
\end{abstract}

\maketitle

\section{INTRODUCTION}

Parton distributions form indispensable ingredients for the analysis 
of all hard-scattering processes involving initial-state hadrons. The 
scale-dependence of these distributions can be derived from 
first principles in terms of an expansion in powers of the strong 
coupling constant~$\as$.
The corresponding $n\,$th-order coefficients governing the evolution are 
referred to as the $n$-loop splitting functions.
Including the terms up to order $\as^{\, n+1}$ in the evolution 
of parton distributions, together with the corresponding results for the 
hard partonic cross sections of a given observable, one obtains the 
N$^{n}$LO (leading-order, next-to-leading-order, 
next-to-next-to-leading-order, etc.) approximation of perturbative QCD.

The standard approximation for most important processes is presently the 
next-to-leading order, the corresponding one- and two-loop splitting
functions being known for a long time.
However, the NNLO corrections need to be included in order to arrive at
quantitatively reliable predictions for hard processes at present and
future high-energy colliders. 

In a series of recent papers~\cite{Moch:2002sn,Moch:2004pa,Vogt:2004mw}, we have 
published the complete unpolarized three-loop splitting functions. 
Here, we present some aspects of the 
methods and selected results.

\section{METHODS}

The method of calculation employs the optical theorem, which relates the 
total cross section for a given process to the imaginary part of the forward 
Compton amplitude.
We study deep-inelastic scattering of a boson with Euclidean (off-shell) momentum $Q$, 
off a parton with (on-shell) momentum $P$ and apply the operator product expansion.
In the Bjorken limit, $Q^2 \to \infty$ and $x=Q^2/(2 P \cdot Q)$ fixed, 
this yields a relation between matrix elements of parton operators of leading twist 
and Mellin moments of Feynman diagrams contributing to the forward Compton amplitude.
Specifically, using dimensional regularization
in $D=4-2\epsilon$, we are able to 
compute the anomalous dimensions $\gamma$ of the parton operators, i.e. the 
integer-$N$ Mellin moments 
of splitting functions $P$, 
\begin{eqnarray}
  \gamma(N) \: = \: 
  - \int_0^1 \!dx\:\, x^{\,N-1}\, P(x) 
\label{eq:anomdim}
\end{eqnarray}
from the divergence in $\epsilon$ of the $N$-th Mellin moment of the corresponding 
Feynman diagrams. 

In general, the (anti-)quark (anti-)quark splitting 
functions, constrained by charge conjugation invariance and flavour 
symmetry, are given for flavours $i,k$ by
\bea
  P_{{\rm q}_{i}{\rm q}_{k}} \: = \: 
  P_{\bar{{\rm q}}_{i}\bar{{\rm q}}_{k}} 
  &\! =\! & \delta_{ik} P_{{\rm q}{\rm q}}^{\,\rm v} 
        + P_{{\rm q}{\rm q}}^{\,\rm s} \, , \\
  P_{{\rm q}_{i}\bar{{\rm q}}_{k}} \: = \: 
  P_{\bar{{\rm q}}_{i}{\rm q}_{k}} 
  &\! =\! & \delta_{ik} P_{{\rm q}\bar{{\rm q}}}^{\,\rm v} 
        + P_{{\rm q}\bar{{\rm q}}}^{\,\rm s} 
  \:\: \nonumber .
\eea
They can be composed into three independent types of non-singlet splitting 
functions,
\beq
\label{eq:ppm} 
  P_{\rm ns}^{\,\pm} \: = \: P_{{\rm q}{\rm q}}^{\,\rm v}
  \pm P_{{\rm q}\bar{{\rm q}}}^{\,\rm v} \:\: ,
\eeq
and 
\beq
\label{eq:pval}
  P_{\rm ns}^{\,\rm v} \: = \: P_{\rm qq}^{\,\rm v} 
  - P_{{\rm q}\bar{{\rm q}}}^{\,\rm v} + n_f (P_{\rm qq}^{\,\rm s} 
  - P_{{\rm q}\bar{{\rm q}}}^{\,\rm s}) \: \equiv \: 
  P_{\rm ns}^{\, -} + P_{\rm ns}^{\,\rm s} \:\: ,
\eeq
which govern the evolution of the quark flavour asymmetries 
$q_{\rm ns}^{\pm}$ and the valence distribution $q_{\rm ns}^{\rm v}$, respectively,
\beq
\label{eq:evolns}
  \frac{d}{d \ln \mu_f^{\,2}} \: q_{\rm ns}^{\: i} \: =\: 
  P_{\rm ns}^{\,i} \,\otimes\, q_{\rm ns}^{\, i}\, ,
\,\,\,\,\,\,\,\,\,\,\,\,\,\,\,\,\, i=\pm,\mbox{v}.
\eeq

The quark-quark singlet splitting function $P_{\rm qq}$ is 
expressed as 
\beq
\label{eq:Pqq}
  P_{\rm qq} \: =\: P_{\rm ns}^{\,+} + n_f (P_{\rm qq}^{\:\rm s}
  + P_{\rm {\bar{q}q}}^{\:\rm s})
  \:\equiv\:  P_{\rm ns}^{\,+} + P_{\rm ps}^{} \:\: ,
\eeq
and gluon-quark and quark-gluon splitting functions are given by
\beq
\label{eq:Poffd}
  P_{\rm qg} \: =\: n_f\, P_{{\rm q}_{i}\rm g} \:\: , \quad
  P_{\rm gq} \: =\: P_{{\rm gq}_{i}}\, ,
\eeq
in terms of the flavour-independent splitting functions $P_{{\rm q}_{i}
\rm g} = P_{\bar{\rm q}_{i}\rm g}$ and $P_{{\rm gq}_{i}} = P_{{\rm g}
\bar{\rm q}_{i}}$.
Thus, the singlet quark distribution $q_{\rm s}^{}$ and the gluon distribution $g$ 
evolve according to 
\beq
\label{eq:evols}
  \frac{d}{d \ln\mu_f^{\,2}}
  \left( \begin{array}{c} \! q_{\rm s}^{} \! \\ g  \end{array} \right) 
  \: = \: \left( \begin{array}{cc} P_{\rm qq} & P_{\rm qg} \\ 
  P_{\rm gq} & P_{\rm gg} \end{array} \right) \otimes 
  \left( \begin{array}{c} \!q_{\rm s}^{}\! \\ g  \end{array} \right) 
  \:\: .
\eeq
Eqs.~(\ref{eq:evolns}) and~(\ref{eq:evols}) form the well-known $2n_f-1$ scalar non-singlet and
$2 \times 2$ singlet evolution equations, where $\otimes$ stands for the Mellin convolution.

In the present calculation, for the complete set of singlet and non-singlet 
splitting functions, we had to compute 18 one-loop Feynman diagrams at LO, 
350 two-loop diagrams at NLO, and a total of 9607 three-loop 
diagrams at NNLO. All diagrams were generated automatically with the diagram 
generator {\sc Qgraf}~\cite{Nogueira:1991ex} and for all symbolic manipulations 
we used the latest version of {\sc Form}~\cite{Vermaseren:2000nd}.
The NNLO calculation required significant enhancements of the capabilities 
of {\sc Form}~\cite{Vermaseren:2002rp}.

In order to illustrate the details of the calculation, let us pick a particular 
Feynman diagram at three loops, which contributes to the forward Compton amplitude 
for the scattering of a photon off a quark.
\begin{center}
\fcolorbox{white}{white}{
\begin{picture}(80,71) (10,-10)
\SetWidth{2.0}
\SetColor{Blue}
\ArrowLine(12,-11)(23,5)
\ArrowLine(77,5)(88,-11)
\Gluon(77,5)(23,5){2.5}{7.71}
\SetColor{Black}
\Text(0,52)[lb]{\large{\Black{$Q$}}}
\Text(0,-14)[lb]{\large{\Black{$P$}}}
\Text(90,52)[lb]{\large{\Black{$Q$}}}
\Text(90,-14)[lb]{\large{\Black{$P$}}}
\SetWidth{0.5}
\Vertex(23,5){1.41}
\Vertex(23,43){1.41}
\GlueArc(23,43)(27,-90,0){2.5}{7.36}
\GlueArc(23,43)(16,-90,0){2.5}{3.86}
\ArrowLine(23,5)(23,16)
\ArrowLine(23,16)(23,27)
\ArrowLine(23,27)(23,43)
\ArrowLine(23,43)(39,43)
\ArrowLine(39,43)(50,43)
\Vertex(23,16){1.41}
\Vertex(23,27){1.41}
\Vertex(39,43){1.41}
\Vertex(50,43){1.41}
\Vertex(77,5){1.41}
\Vertex(77,43){1.41}
\ArrowLine(50,43)(77,43)
\ArrowLine(77,43)(77,5)
\Photon(77,43)(88,59){2.5}{3}
\Photon(23,43)(12,59){2.5}{3}
\end{picture}
}
\end{center}
Here and below, the fat line indicates the flow of the quark momentum $P$ through the diagram. 

In a first step, we need a classification of the topology of the loop integral. 
This is done in two steps. Any loop integral with external momenta $P$ and $Q$, we classify according to
\begin{enumerate}
\item the topology of the underlying ``two-point function'' if we nullify $P$.
\item the flow of the parton momentum $P$.
\end{enumerate}
The first step yields at the top-level the topology types {\it ladder}, {\it benz} 
and {\it non-planar}, where we use the notation of Refs.~\cite{Gorishnii:1989gt,Larin:1991fz}.
The second step distinguishes, within each topology, between so-called basic building blocks (BBB) 
which have a simple $P$-flow with one $P$-dependent propagator only, and 
so-called composite building blocks (CBB), which have a more complicated $P$-flow.
At the top-level, there are 10 BBB (3 ladder, 5 benz and 2 non-planar) 
and 32 CBB (10 ladder, 16 benz and 6 non-planar). 
The smaller number of non-planar topologies is due to 
symmetries. 
For illustration purposes, we represent the complete set of top-level BBB by pictograms as follows.
\begin{center}
\fcolorbox{white}{white}{
\begin{picture}(167,141) (45,-30)
\SetWidth{0.5}
\SetColor{Black}
\CArc(61,100)(11,90,270)
\Line(61,111)(61,89)
\Line(45,100)(50,100)
\CArc(77,100)(11,-90,90)
\Line(88,100)(93,100)
\Line(61,111)(77,111)
\Line(61,89)(77,89)
\Line(77,111)(77,89)
\SetWidth{2.0}
\SetColor{Blue}
\CArc(61,100)(11,90,135)
\SetWidth{0.5}
\SetColor{Black}
\Line(121,111)(137,111)
\Line(104,100)(109,100)
\CArc(121,100)(11,90,270)
\CArc(137,100)(11,-90,90)
\Line(148,100)(153,100)
\Line(121,89)(137,89)
\Line(121,111)(121,89)
\Line(137,111)(137,89)
\SetWidth{2.0}
\SetColor{Blue}
\Line(121,111)(129,111)
\SetWidth{0.5}
\SetColor{Black}
\Line(180,111)(180,89)
\Line(164,100)(169,100)
\CArc(180,100)(11,90,270)
\CArc(196,100)(11,-90,90)
\Line(207,100)(212,100)
\Line(180,111)(196,111)
\Line(180,89)(196,89)
\Line(196,111)(196,89)
\SetWidth{2.0}
\SetColor{Blue}
\Line(180,111)(180,100)
\SetWidth{0.5}
\SetColor{Black}
\CArc(68,62)(17.03,87,447)
\Line(67,62)(67,46)
\Line(56,73)(67,62)
\Line(79,74)(67,62)
\SetWidth{2.0}
\SetColor{Blue}
\CArc(65,62)(14.5,130,160)
\SetWidth{0.5}
\SetColor{Black}
\Line(46,62)(51,62)
\Line(85,62)(90,62)
\Line(56,30)(67,19)
\SetWidth{2.0}
\SetColor{Blue}
\Line(55.5,30.5)(62,24)
\SetWidth{0.5}
\SetColor{Black}
\CArc(68,19)(17.03,87,447)
\Line(46,19)(51,19)
\Line(85,19)(90,19)
\Line(67,19)(67,3)
\Line(79.5,31.5)(67,19)
\Line(121,19)(121,3)
\CArc(122,19)(17.03,87,447)
\Line(100,19)(105,19)
\Line(139,19)(144,19)
\Line(110,30)(121,19)
\Line(132,30)(121,19)
\SetWidth{2.0}
\SetColor{Blue}
\Line(121,19)(121,11)
\SetWidth{0.5}
\SetColor{Black}
\CArc(122,62)(17.03,87,447)
\Line(121,62)(121,46)
\Line(110,73)(121,62)
\Line(133,74)(121,62)
\SetWidth{2.0}
\SetColor{Blue}
\CArc(121,63)(15.5,90,138)
\SetWidth{0.5}
\SetColor{Black}
\Line(100,62)(105,62)
\Line(139,62)(144,62)
\CArc(176,62)(17.03,87,447)
\Line(175,62)(175,46)
\Line(164,73)(175,62)
\Line(187,74)(175,62)
\Line(154,62)(159,62)
\Line(193,62)(198,62)
\SetWidth{2.0}
\SetColor{Blue}
\CArc(174,62)(15.5,-180,-135)
\SetWidth{0.5}
\SetColor{Black}
\CArc(61,-19)(11,90,270)
\Line(45,-19)(50,-19)
\CArc(77,-19)(11,-90,90)
\Line(88,-19)(93,-19)
\Line(61,-8)(77,-8)
\Line(61,-8)(77,-30)
\Line(77,-8)(61,-30)
\Line(61,-30)(77,-30)
\SetWidth{2.0}
\SetColor{Blue}
\CArc(61,-19)(11,90,135)
\SetWidth{0.5}
\SetColor{Black}
\Line(121,-8)(137,-8)
\Line(104,-19)(109,-19)
\CArc(121,-19)(11,90,270)
\CArc(137,-19)(11,-90,90)
\Line(148,-19)(153,-19)
\Line(121,-30)(137,-30)
\Line(121,-8)(137,-30)
\Line(137,-8)(121,-30)
\SetWidth{2.0}
\SetColor{Blue}
\Line(121,-8)(129,-8)
\end{picture}
}
\end{center}

A quick inspection shows, that our example is a particular ladder type BBB,
\begin{flushleft}
\fcolorbox{white}{white}{
\begin{picture}(86,22) (45,-30)
\SetWidth{0.5}
\SetColor{Black}
\CArc(61,-19)(11,90,270)
\Line(61,-8)(61,-30)
\Line(45,-19)(50,-19)
\CArc(77,-19)(11,-90,90)
\Line(88,-19)(93,-19)
\Line(61,-8)(77,-8)
\Line(61,-30)(77,-30)
\Line(77,-8)(77,-30)
\SetWidth{2.0}
\SetColor{Blue}
\CArc(61,-19)(11,90,135)
\Text(98,-33)[lb]{
\Black{$\displaystyle {\Large{= \frac{ \left( 2\, P \cdot Q \right)^N}{\left(Q^2 \right)^{N+\alpha}}}}\,\, 
    {\large{C_N\,.}}$}
}
\end{picture}
}
\end{flushleft}

The equation indicates that we calculate the $N$-th Mellin moment of this diagram.
This is  precisely the dimensionless coefficient $C_N$ 
given on the right hand side.
On the left hand side, the fat line in the pictogram represents the flow of the momentum $P$,
in form of only one internal propagator containing both, a loop momentum, say $l_1$, and the 
external quark momentum $P$. Thus, on the left hand side the $N$-th term in a Taylor expansion 
generates the contribution to the $N$-th Mellin moment, 
\begin{eqnarray}
{1 \over (P - l_1)^2}
= \sum\limits_i {(2 P\cdot l_1)^i \over (l_1^2)^{i+1}}
\longrightarrow 
{(2 P\cdot l_1)^N \over (l_1^2)^N} \, .
\end{eqnarray}

Thus far, the set-up is completely analogous to the calculation of 
the lowest six/seven (even or odd) integer-$N$ Mellin moments 
of the three-loop splitting functions~\cite{Larin:1994vu,Larin:1997wd,Retey:2000nq}, 
where the {\sc Mincer} program~\cite{Gorishnii:1989gt,Larin:1991fz} was used as a tool to solve the integrals. 
As a new feature here we are dealing with symbolic $N$. This was tested before up to two loops~\cite{Moch:1999eb}.
At three loops, it leads, for instance, to integrals of the type 
\begin{flushleft}
\fcolorbox{white}{white}{
\begin{picture}(86,22) (45,-30)
\SetWidth{2.0}
\Text(40,-10)[lb]{\Blue{${\large{n,k}}$}}
\SetWidth{0.5}
\SetColor{Black}
\CArc(61,-19)(11,90,270)
\Line(61,-8)(61,-30)
\Line(45,-19)(50,-19)
\CArc(77,-19)(11,-90,90)
\Line(88,-19)(93,-19)
\Line(61,-8)(77,-8)
\Line(61,-30)(77,-30)
\Line(77,-8)(77,-30)
\Text(98,-34)[lb]{\Black{$\displaystyle {\large{ {\displaystyle
=  \int\, \prod_n^3\,} d^Dl_n\,}} {\Large{
{(2 P \cdot l_1)^k \over (l_1^2)^{n}}
\, \frac{1}{l_2^2\, \dots l_8^2}\, ,}}$}}
\end{picture}
}
\end{flushleft}
where the powers $n,k$ are symbolic.

However, we can switch to non-symbolic (fixed) positive powers $n,k$ and values of $N$ 
at any point of the derivations and calculations, 
after which the {\sc Mincer} program can be invoked to verify that the results are correct. 
From a practical point of view this is the most powerful feature of the Mellin-space approach, 
as it allows for extremely efficient checks.

Having classified the integrals, we actually need to solve them, too. 
To this end, let us start the discussion with the most general form of a three-loop integral 
in one of the top-level topologies ladder, benz or non-planar,
\begin{eqnarray}
\lefteqn{
I(N,{\vec{\mu}},{\vec{\nu}},{\vec{\kappa}}) = \int\, \prod_{n=1}^3\, d^Dl_n\,} \\
&\!\!\!\displaystyle
\frac{(2P\cdot l_1)^{\kappa_1}\, \dots (2P\cdot l_8)^{\kappa_8} 
(2 l_9 \cdot l_{10})^{\kappa_9}}{(l_1^2)^{\mu_1}\,((P-l_1)^2)^{\nu_1}\, \dots (l_8^2)^{\mu_8}\,((P-l_8)^2)^{\nu_8}}\, .
\nonumber
\end{eqnarray}
Here, ${\vec{\mu}}=\mu_1,\dots,\mu_8$,
${\vec{\nu}}=\nu_1,\dots,\nu_8$ and 
${\vec{\kappa}}=\kappa_1,\dots,\kappa_9$ are symbolic parameters.
Furthermore, the $l_i$ are for $i=1,2,3$ independent loop momenta. For $i=4,\dots,8$ they can be expressed 
by the former and the external momentum $Q$, the precise relations depending on the topology.
The momenta $l_9, l_{10}$ denote an irreducible scalar product between the momenta $Q$ and/or $l_i$ ,$i=1,\dots,8$, 
again the precise relation being topology depended.

Applying relations based on integration 
by parts~\cite{'tHooft:1972fi,Tkachov:1981wb,Chetyrkin:1981qh,Tkachov:1984xk}, 
scaling equations, form-factor analysis~\cite{Passarino:1979jh} and some 
equations~\cite{Moch:2004pa} that fall in a special category because they involve higher twist and 
a careful study of the parton-momentum limit $P\cdot P \rightarrow 0$,
we arrive at a system of linear equations for a given integral 
$I(N,{\vec{\mu}},{\vec{\nu}},{\vec{\kappa}})$ under consideration.

Solving this linear system amounts to finding 
a scheme in which $I(N,{\vec{\mu}},{\vec{\nu}},{\vec{\kappa}})$ is 
mapped to a set of master integrals and integrals of simpler topologies.   
The general strategy applies two basic rules of mapping,
\begin{enumerate}
  \item non-planar $\longrightarrow$ benz $\longrightarrow$ ladder.
  \item CBB $\longrightarrow$ BBB.
\end{enumerate}
Here, the first rule is understood to hold for common sub-topologies.
For instance, the non-planar topology contains common sub-topologies with 
the benz and the ladder topology~\cite{Gorishnii:1989gt,Larin:1991fz}.

In an operator approach, this can be realized by 
diagonalizing the linear system  for symbolic indices. 
One thus obtains lowering operators for individual $\nu_i$ or $\mu_i$, 
as well as recursion relations in the Mellin moment $N$.
The latter constitute difference equations, which may generally be written as 
\begin{eqnarray}
\label{eq:diffeq}
\lefteqn{
  a_0(N)\, I(N) + a_1(N)\,  I(N-1)} \\ 
& &+ \ldots + a_m(N)\,  I(N-m) \: = \: G(N)\, .
\nonumber
\end{eqnarray}

To illustrate the latter, let us give an extremely simple example 
for a single-step difference equation in $N$, which occurs 
in the reduction of a particular type of CBB,
\begin{center}
\fcolorbox{white}{white}{
\begin{picture}(198,56) (15,-30)
\SetWidth{0.5}
\SetColor{Black}
\CArc(31,15)(11,90,270)
\Line(31,26)(31,4)
\Line(15,15)(20,15)
\CArc(47,15)(11,-90,90)
\Line(58,15)(63,15)
\SetWidth{2.0}
\SetColor{Blue}
\Line(31,26)(47,26)
\SetWidth{0.5}
\SetColor{Black}
\Line(31,4)(47,4)
\Line(47,26)(47,4)
\SetWidth{2.0}
\SetColor{Blue}
\CArc(31,15)(11,90,135)
\CArc(46,15)(11,45,90)
\SetWidth{0.5}
\SetColor{Black}
\CArc(107,-19)(11,90,270)
\Line(107,-8)(107,-30)
\Line(91,-19)(96,-19)
\CArc(123,-19)(11,-90,90)
\Line(134,-19)(139,-19)
\SetWidth{2.0}
\SetColor{Blue}
\Line(107,-8)(123,-8)
\SetWidth{0.5}
\SetColor{Black}
\Line(107,-30)(123,-30)
\Line(123,-8)(123,-30)
\Vertex(131,-12){1.41}
\SetWidth{2.0}
\SetColor{Blue}
\CArc(107,-19)(11,90,135)
\Text(52,-29)[lb]{\Large{\Black{$+{2 \over N+2}$}}}
\SetWidth{0.5}
\SetColor{Black}
\CArc(181,15)(11,90,270)
\Line(181,26)(181,4)
\Line(165,15)(170,15)
\CArc(197,15)(11,-90,90)
\Line(208,15)(213,15)
\SetWidth{2.0}
\SetColor{Blue}
\Line(181,26)(197,26)
\SetWidth{0.5}
\SetColor{Black}
\Line(181,4)(197,4)
\Line(197,26)(197,4)
\SetWidth{2.0}
\SetColor{Blue}
\CArc(181,15)(11,90,135)
\CArc(196,15)(11,45,90)
\Text(66,5)[lb]{\Large{\Black{$= - {N+3+3\epsilon \over N+2}\, {2P\cdot Q \over Q^2}$}}}
\end{picture}
}
\end{center}
In the pictogram, all lines denote propagators of unit power, except the one with a blob,
which has to be taken to second power. Again, it is understood that the equation holds for 
the $N$-th Mellin moment of the diagrams and the fat lines indicate propagators with the 
momentum $P$.

Employing the notation of Eq.~(\ref{eq:diffeq}), we can write 
the single-step difference equations as 
\begin{eqnarray}
\label{eq:examplesinglestep}
I(N) = - \frac{N\!+\!3\!+\!3\epsilon}{N\!+\!2}\,  I(N-1)
 + \frac{2}{N\!+\!2}\, G(N).
\end{eqnarray}

As a remark on the side, imagine for a moment, the function $G(N)$ 
on right hand side would be multiplied by an additional factor $\epsilon^{-1}$. 
If present, such a so-called spurious pole in $\epsilon$ would make 
Eq.~(\ref{eq:examplesinglestep}) useless. It would ruin the accuracy 
of the expansion when working only a to given cutoff in powers of $\epsilon$. 
(We do so both for reasons of economy and because we cannot evaluate some integrals 
easily beyond certain powers in $\epsilon$.) Thus, spurious poles have to be avoided 
and one of the greatest difficulties in deriving reduction schemes is to 
indeed avoid them.

Eq.~(\ref{eq:examplesinglestep}) can be solved in closed form,
\begin{eqnarray}
\label{eq:examplediff}
\lefteqn{
I(N) =  (-1)^N\, 
        \frac{\prod_{j=1}^N (j\!+\!3\!+\!3\epsilon)}{\prod_{j=1}^N (j\!+\!2)} I(0) }\\
&&
                +(-1)^N\,\sum_{i=1}^N\, (-1)^j\,\frac{\prod_{j=i+1}^N (j\!+\!3\!+\!3\epsilon)}{\prod_{j=i}^N (j\!+\!2)}
		G(i)\, .
\nonumber
\end{eqnarray}

Eq.~(\ref{eq:examplediff}) is an example for the occurrence of nested sums in the 
calculation. 
Its solution requires as input the boundary value $I(0)$, 
which can be obtained with the {\sc Mincer} program. 
The inhomogeneous term $G(N)$ is assumed to be already known. 
It has to be calculated by similar means, i.e. reductions and recursions. 
The successive way of solving difference equations induces 
a strict hierarchy for all topology classes in the reduction scheme.

As a matter of fact, Eq.~(\ref{eq:examplediff}) is a special case of a general recursion which was 
derived for $I(N,{\vec{\mu}},{\vec{\nu}},{\vec{\kappa}})$ with symbolic indices, like most other 
recursions as well. 
As such, they allow for an efficient implementation in {\sc Form} resulting in a largely 
automatic build-up of nested sums.

The solution of nested sums as in Eq.~(\ref{eq:examplediff}) results in 
harmonic sums~\cite{Gonzalez-Arroyo:1979df,Gonzalez-Arroyo:1980he,Vermaseren:1998uu,Blumlein:1998if,Moch:2001zr}
which are recursively defined as 
\begin{eqnarray}
\label{eq:def-S}
S_{\pm m_1,...,m_k}(N) = 
        \sum\limits_{i=1}^N  \frac{(\pm 1)^i}{i^{m_1}} S_{m_2,...,m_k}(i)\, .
\end{eqnarray}

In particular, we use four main 
algorithms for harmonic sums, which rely on the underlying algebra~\cite{Hoffman1,Hoffman2}.
The algorithms express products or sums of nested sums again in the basis Eq.(\ref{eq:def-S}) 
of harmonic sums.
Specifically, they act on products,
\begin{eqnarray}
S_{m_1,...,m_k}(N)\, S_{n_1,...,n_l}(N)\, ,
\end{eqnarray}
sums involving $j$ and $N-j$ 
\begin{eqnarray}
{\lefteqn{
\sum\limits_{j=1}^{N-1}\, 
{1 \over j^{m_1}} S_{m_2,...,m_k}(j)\,
{1 \over (N-j)^{n_1}} }}
\\&&
\times S_{n_2,...,n_l}(N-j)\, ,
\nonumber
\end{eqnarray}
conjugations 
\begin{eqnarray}
- \sum\limits_{j=1}^{N}\, 
\left(\!\!\!\! \begin{array}{c} N \\
  j \end{array} \!\!\!\!\right)\, (-1)^j \, 
{1 \over j^{m_1}} S_{m_2,...,m_k}(j)\, ,
\end{eqnarray}
and on sums involving binomials, $j$ and $N-j$, 
\begin{eqnarray}
{\lefteqn{
- \sum\limits_{j=1}^{N-1}\, \left(\!\!\!\! \begin{array}{c} N \\
  j \end{array} \!\!\!\!\right)\, (-1)^j\,
{1 \over j^{m_1}} S_{m_2,...,m_k}(j)
}}
\\&&\nonumber
\times 
{1 \over (N-j)^{n_1}}   S_{n_2,...,n_l}(N-j)\, .
\end{eqnarray}

The solution to Eq.~(\ref{eq:examplediff}) up to order $\epsilon^{-1}$ as required 
for the calculation of the splitting functions can be obtained,
\begin{eqnarray}
\lefteqn{
I(N) = (-1)^N \* {1 \over \epsilon^2}  \*  \biggl( 
  {4 \over 3} \* {S_{1}(N\!+\!1) \over N\!+\!1}
+ {8 \over 3} \* {S_{1}(N\!+\!1) \over (N\!+\!1)^2} 
 }
\label{eq:exampleresult}
\\
&&\!\!\!\!\!\! 
\nonumber
+ {4 \over 3} \* {S_{1}(N\!+\!2) \over N\!+\!2} 
+ {4 \over 3} \* {S_{1}(N\!+\!2) \over (N\!+\!2)^2} 
+ {4 \over 3} \* S_{1}(N) 
\\&&\!\!\!\!\!\! 
\nonumber
+ {2 \over 3} \* S_{1,2}(N) 
+ {2 \over 3} \* {S_{2}(N\!+\!1)  \over N\!+\!1} 
+ {2 \over 3} \* {S_{2}(N\!+\!2)  \over N\!+\!2}
\\&&\!\!\!\!\!\! 
\nonumber
- 2 \* S_{2}(N) 
- {4 \over 3} \* N \* S_{2}(N)  
+ 4 \* S_{2,1}(N) 
\\&&\!\!\!\!\!\! 
\nonumber
+ {4 \over 3}  \* N \* S_{2,1}(N)
- 6 \* S_{3}(N) 
- 2  \* N \* S_{3}(N)
\\&&\!\!\!\!\!\! 
\nonumber
- {8 \over 3} \* {1 \over (N\!+\!1)^2}
- 4 \* {1 \over (N\!+\!1)^3} 
- {4 \over 3} \* {1 \over (N\!+\!2)^2} 
\\&&\!\!\!\!\!\! 
\nonumber
- 2 \* {1 \over (N\!+\!2)^3} \biggr)
+ (-1)^N \* {1 \over \epsilon}  \*  \biggl(  
- 16 \* {S_{1}(N\!+\!1) \over N\!+\!1}
\\&&\!\!\!\!\!\! 
\nonumber
- {88 \over 3} \* {S_{1}(N\!+\!1) \over (N\!+\!1)^2}
- {20 \over 3} \* {S_{1}(N\!+\!1) \over (N\!+\!1)^3} 
- 16 \* {S_{1}(N\!+\!2) \over N\!+\!2} 
\\&&\!\!\!\!\!\! 
\nonumber
- {44 \over 3} \* {S_{1}(N\!+\!2) \over (N\!+\!2)^2} 
- {10 \over 3} \* {S_{1}(N\!+\!2) \over (N\!+\!2)^3} 
- 20 \* S_{1}(N) 
\\&&\!\!\!\!\!\! 
\nonumber
+ {8 \over 3} \* {S_{1,1}(N\!+\!1) \over N\!+\!1} 
+ {8 \over 3} \* {S_{1,1}(N\!+\!1) \over (N\!+\!1)^2} 
+ {8 \over 3} \* {S_{1,1}(N\!+\!2) \over N\!+\!2} 
\\&&\!\!\!\!\!\! 
\nonumber
+ {8 \over 3} \* S_{1,1}(N) 
+ {10 \over 3} \* S_{1,1,2}(N) 
+ {10 \over 3} \* {S_{1,2}(N\!+\!1) \over N\!+\!1}
\\&&\!\!\!\!\!\! 
\nonumber
+ {10 \over 3} \* {S_{1,2}(N\!+\!2) \over N\!+\!2}
- 16 \* S_{1,2}(N) 
- 4 \* N \* S_{1,2}(N)
\\&&\!\!\!\!\!\! 
\nonumber
+ 14 \* S_{1,2,1}(N) 
+ 4 \* N \* S_{1,2,1}(N)
- 24 \* S_{1,3}(N) 
\\&&\!\!\!\!\!\! 
\nonumber
- 6 \* N \* S_{1,3}(N)
- {58 \over 3} \* {S_{2}(N\!+\!1) \over N\!+\!1}
- {40 \over 3} \* {S_{2}(N\!+\!1) \over (N\!+\!1)^2} 
\\&&\!\!\!\!\!\! 
\nonumber
- {46 \over 3} \* {S_{2}(N\!+\!2) \over N\!+\!2} 
- 6 \* {S_{2}(N\!+\!2) \over (N\!+\!2)^2} 
+ {56 \over 3} \* S_{2}(N) 
\\&&\!\!\!\!\!\! 
\nonumber
+ 20 \* N \* S_{2}(N)
+ 10 \* {S_{2,1}(N\!+\!1) \over N\!+\!1} 
+ 6 \* {S_{2,1}(N\!+\!2) \over N\!+\!2} 
\\&&\!\!\!\!\!\! 
\nonumber
- {134 \over 3} \* S_{2,1}(N) 
- {56 \over 3}  \* N \* S_{2,1}(N)
+ {16 \over 3} \* S_{2,1,1}(N) 
\\&&\!\!\!\!\!\! 
\nonumber
+ {8 \over 3}  \* N \* S_{2,1,1}(N)
- {62 \over 3} \* S_{2,2}(N) 
- {22 \over 3}  \* N \* S_{2,2}(N)
\\&&\!\!\!\!\!\! 
\nonumber
- 18 \* {S_{3}(N\!+\!1) \over N\!+\!1}
- 12 \* {S_{3}(N\!+\!2) \over N\!+\!2} 
+ 76 \* S_{3}(N) 
\\&&\!\!\!\!\!\! 
\nonumber
+ {100 \over 3}  \* N \* S_{3}(N)
- 10 \* S_{3,1}(N) 
- {10 \over 3}  \* N \* S_{3,1}(N)
\\&&\!\!\!\!\!\! 
\nonumber
+ 36 \* S_{4}(N) 
+ 12  \* N \* S_{4}(N)
+ 32 \* {1 \over (N\!+\!1)^2} 
\\&&\!\!\!\!\!\! 
\nonumber
+ {164 \over 3} \* {1 \over (N\!+\!1)^3} 
+ 24 \* {1 \over (N\!+\!1)^4}
+ 16 \* {1 \over (N\!+\!2)^2} 
\\&&\!\!\!\!\!\! 
\nonumber
+ {82 \over 3} \* {1 \over (N\!+\!2)^3} 
+ 12 \* {1 \over (N\!+\!2)^4} \biggr)\, .
\end{eqnarray}

The result in Eq.~(\ref{eq:exampleresult}) is neither short nor inexpensive 
to calculate. Moreover, each integral is typically used many times.
Therefore, to save computer time and disk space, we have tabulated large numbers of CBB and BBB integrals.
Only this tabulation rendered the calculation feasible with current computing resources. 
For the complete project we have collected table-bases with more than 
$100.000$ integrals and a total size of tables of more than 3 GBytes.

Subsequently, the database of integrals was used for the calculation of all Feynman diagrams yielding 
the unrenormalized results in Mellin-space in terms of the invariants determined 
by the colour group~\cite{vanRitbergen:1998pn}, harmonic sums and the values 
$\zeta_3$, $\zeta_4$, $\zeta_5$ of the Riemann \mbox{$\zeta$-function}. 
In physics results the terms with $\zeta_4$ cancel in $N$-space. 
The renormalization was carried out in the \MSb-scheme~\cite{'tHooft:1973mm,Bardeen:1978yd} 
and the procedure is again the same as for the fixed-moment 
calculations~\cite{Larin:1994vu,Larin:1997wd,Retey:2000nq}.

\section{RESULTS}

Now we present the anomalous dimensions $\gamma(\as,N)$ 
up to the third order in the running coupling constant 
$\as$,  the expansion coefficients being normalized as
\beq
 \gamma\left(\as,N\right) \: = \: \sum_{n=0}\,
  \left(\frac{\as}{4\pi}\right)^{n+1} \gamma^{\,(n)}(N)
  \:\: .
\eeq
Our analytical results have been presented in Refs.~\cite{Moch:2002sn,Moch:2004pa,Vogt:2004mw} 
and are too long to be reproduced here.
We agree with all partial results available
in the literature, in particular we reproduce the lowest six even integer singlet and 
seven even/odd integer non-singlet moments\footnote{The moment $N=16$ of $\gamma^{\,(2)+}_{\,\rm ns}$ 
has recently been computed as an additional check~\cite{Blumlein:2004bb}.} 
computed before \cite{Larin:1994vu,Larin:1997wd,Retey:2000nq}.
\begin{figure}[htb]
\begin{center}
\includegraphics[width=7.5cm]{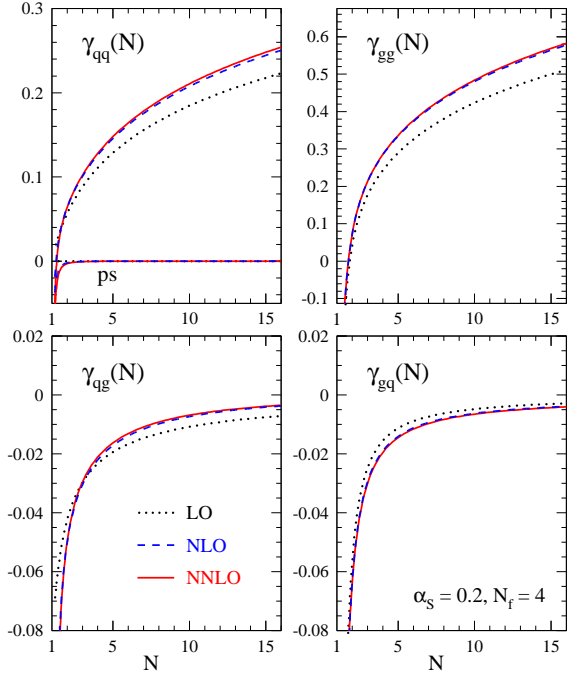}
\end{center}
\vspace*{-9mm}
\caption{The perturbative expansion of the singlet anomalous dimensions $\gamma(\alpha_s,N)$.}
\label{fig:1}
\end{figure}
\begin{figure}[htb]
\begin{center}
\includegraphics[width=7.5cm]{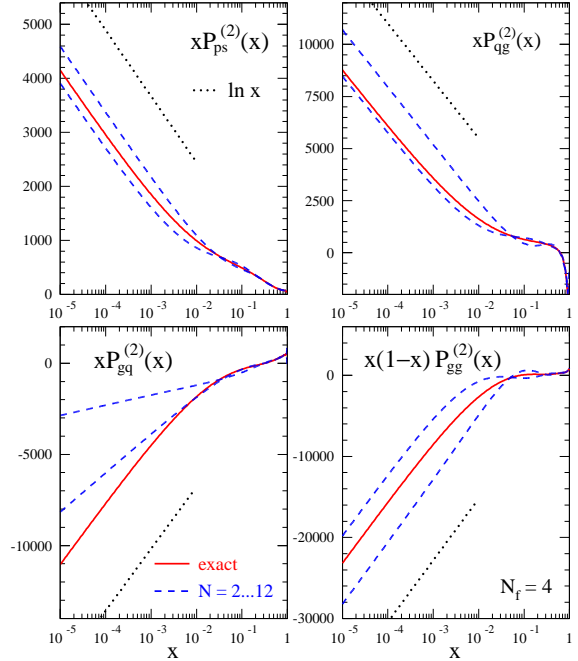}
\end{center}
\vspace*{-9mm}
\caption{The three-loop singlet splitting functions 
$P^{\,(2)}_{\rm ab}$ with the leading small-$x$ terms (dotted) 
and the fixed-moment estimates (dashed).
}
\label{fig:2}
\end{figure}
\begin{figure}[htb]
\begin{center}
\includegraphics[width=7.5cm]{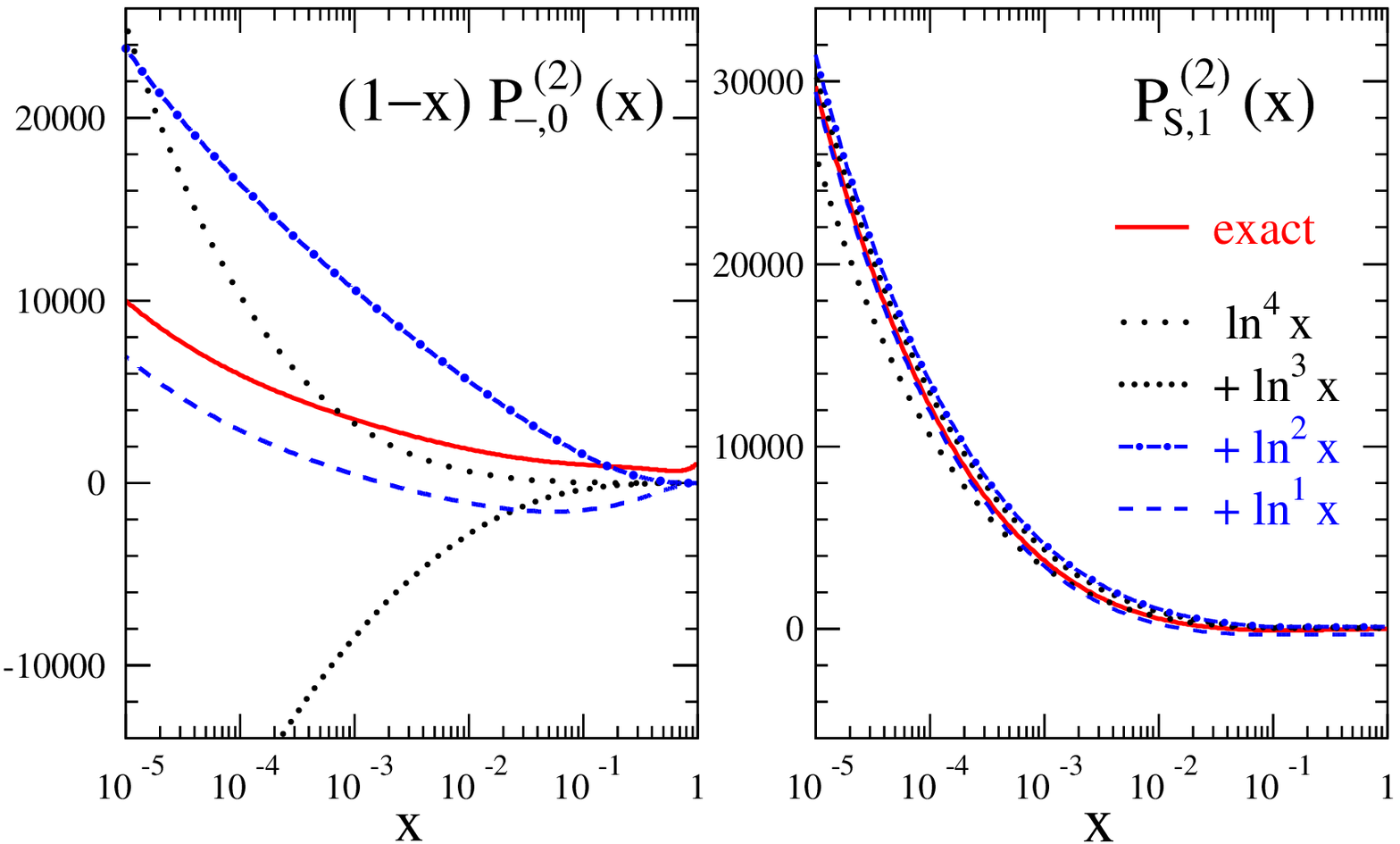}
\end{center}
\vspace*{-9mm}
\caption{The non-singlet three-loop splitting functions $P_{\rm ns}^{\, -}$ ($n_f$-independent part) 
and $P_{\rm ns}^{\: s}/n_f$.}
\label{fig:3}
\end{figure}

The numerical results for the singlet anomalous dimension, 
i.e. the Mellin transforms of the matrix entries in Eq.~(\ref{eq:evols}) are illustrated in Fig.~\ref{fig:1}.
In the top row of Fig.~\ref{fig:1}, we show the perturbative expansion of the diagonal anomalous 
 dimensions $\gamma_{\,\rm qq}(N)$ and $\gamma_{\,\rm gg}(N)$ for four 
 flavours at $\as = 0.2$. The pure-singlet (ps) contribution to 
 $\gamma_{\,\rm qq}$ as defined in Eq.~(\ref{eq:Pqq}) is displayed separately. 
The bottom row of Fig.~\ref{fig:1} shows the off-diagonal anomalous dimensions
 $\gamma_{\,\rm qg}(N)$ and $\gamma_{\,\rm gq}(N)$.
For all cases, the NNLO corrections are significantly smaller than the NLO contributions 
and amount to less than 2\% and 1\% for the large diagonal quantities $\gamma_{\,\rm qq}$ and 
$\gamma_{\,\rm gg}$, respectively, for $N>2$.

In Bjorken $x$-space, the N$^{n}$LO splitting functions $P^{\,(n)}(x)$ in
\beq
\label{eq:Pexp}
 P\left(\as,x\right) \: = \: \sum_{n=0}\,
  \left(\frac{\as}{4\pi}\right)^{n+1} P^{\,(n)}(x)\, ,
\eeq
can be obtained from Eq.~(\ref{eq:anomdim}) 
and expressed in terms of harmonic polylogarithms~\cite{Goncharov,Borwein,Remiddi:1999ew} 
by means of an inverse Mellin transformation. 
This is a completely algebraic procedure based on the fact that harmonic sums occur 
as coefficients of the Taylor expansion of harmonic polylogarithms.

Again, the analytical results have been given in Refs.~\cite{Moch:2002sn,Moch:2004pa,Vogt:2004mw}. 
They agree with the available resummation predictions~\cite{Catani:1994sq,%
Fadin:1998py,Kirschner:1983di,Blumlein:1996jp} for the leading small-$x$ logarithms, 
and those for large-$n_f$ results \cite{Gracey:1994nn,Bennett:1997ch}.
In addition, we have also provided easy-to-use accurate
parameterizations. 

Our results respect the supersymmetric relation between all four singlet splitting functions for
$\,C_A = C_F = n_f\, $ to the extend expected for the \MSb\ scheme. 
At large $x$ we agree with Refs.~\cite{Korchemsky:1989si,Berger:2002sv} and 
determine for $P_{\rm ns}^{\,\pm,\rm v}$, $P_{\rm qq}$ and $P_{\rm gg}$ 
the coefficients of the leading $1/(1-x)_+$ terms. We find that the coefficients of the leading 
integrable term $\ln (1-x)$ at order $n=2,\,3$ are proportional to the 
coefficient of the $+$-distribution $1/(1-x)_+$ at order $n-1$, a result that
seems to point to a yet unexplored structure.
Furthermore, we verify the expected simple relation between the leading $1/(1-x)_+$ terms 
of $P_{\rm qq}$ and $P_{\rm gg}$.

In Fig.~\ref{fig:2} we plot the three-loop singlet splitting functions of Eq.~(\ref{eq:evols}) 
in the small-$x$ region, where the leading small-$x$ behaviour is of the type $x^{\,-1} \ln x$.
This agrees with the prediction of the leading logarithmic 
BFKL equation~\cite{Kuraev:1977fs,Balitsky:1978ic,Jaroszewicz:1982gr}. 
In the top row of Fig.~\ref{fig:2}, we give the three-loop splitting functions 
$P^{\,(2)}_{\rm ps}$ (pure-singlet quark-quark) and 
$P^{\,(2)}_{\rm qg}$ (gluon-quark) for four flavours, 
multiplied by $x$ for display purposes. Also shown
 are the uncertainty band derived in Ref.~\cite{vanNeerven:2000wp}
 using the lowest six even-integer moments
 \cite{Larin:1997wd,Retey:2000nq} and the leading small-$x$ terms
 \cite{Catani:1994sq}.
In the bottom row of Fig.~\ref{fig:2}, we show the three-loop splitting functions 
$P^{\,(2)}_{\rm gq}$ (quark-gluon) and 
$P^{\,(2)}_{\rm gg}$ (gluon-gluon). 
For $P^{\,(2)}_{\rm gq}$, the leading small-$x$ 
 contribution was unknown before the present calculation, 
for $P^{\,(2)}_{\rm gg}$ the leading small-$x$ term has been first 
 obtained in Ref.~\cite{Fadin:1998py}.
$P^{\,(2)}_{\rm gg}$ as a diagonal quantity has 
 been additionally multiplied by $(1-x)$.
As illustrated in Fig.~\ref{fig:2} the leading small-$x$ terms alone do not provide good approximations 
of the full results at experimentally relevant small values of~$x$. 
At $x=10^{-4}$, for example, they exceed the exact values of $P_{\rm ab}^{(2)}(x)$ 
by factors between 1.6 and 2.0 for $n_f=4$. 

In Fig.~\ref{fig:3} we display the three-loop non-singlet splitting functions 
$P_{-}^{\,(2)}(x)$ and  $P^{\,(2)\rm s}_{\,\rm ns}$.
 On the left hand side of Fig.~\ref{fig:3}, the $n_f$-independent three-loop contribution 
 $P_{-,0}^{\,(2)}(x)$ to the splitting function $P_{\rm ns}^{\, -}(x)$, 
 multiplied by $(1-x)$ for display purposes is shown.
 Also shown is a comparison of our
 exact result to the small-$x$ expansion in powers of $\ln x$.
Here, the leading small-$x$ terms of the type $\ln^4 x$ were known before for 
$P_{\rm ns}^{\,(2)\pm}(x)$~\cite{Kirschner:1983di,Blumlein:1996jp}.
As can be seen from Fig.~\ref{fig:3} on the left, 
the coefficients of $\ln^{\, k}x$ increase
sharply with decreasing power $k$. 
Including all logarithmically enhanced terms, one still underestimates the complete result  
by a factor as large as 2.0 for $P^{\,(2)-}_{\,\rm ns}$ at $x = 10^{-4}$.

The three-loop contribution $P^{\,(2)\rm s}_{\,\rm ns}$ exhibits a new 
colour structure $d^{abc}d_{abc}/n_c$ which appears for the first time at three loops. 
It is due to Feynman graphs of the following type, involving axial currents 
(the quarks couple to $W$-bosons),
\begin{center}
\fcolorbox{white}{white}{
\begin{picture}(88,87) (13,-9)
\SetWidth{0.5}
\SetColor{Black}
\Vertex(27,33){1.41}
\Vertex(87,33){1.41}
\Vertex(57,33){1.41}
\Vertex(27,8){1.41}
\Vertex(57,8){1.41}
\Vertex(87,8){1.41}
\ArrowLine(27,8)(57,8)
\ArrowLine(57,8)(87,8)
\ArrowLine(87,33)(57,33)
\ArrowLine(57,33)(27,33)
\Gluon(27,33)(27,7){2}{5.14}
\Gluon(57,33)(57,7){2}{5.14}
\Gluon(87,33)(87,7){2}{5.14}
\Vertex(27,60){1.41}
\Photon(87,58)(100,76){2}{4}
\Vertex(87,60){1.41}
\ArrowLine(87,60)(87,34)
\ArrowLine(27,60)(88,60)
\ArrowLine(27,34)(27,60)
\ArrowLine(14,-10)(27,8)
\ArrowLine(87,8)(100,-10)
\Photon(14,76)(25,61){2}{3}
\end{picture}
}
\end{center}
Recall that at one and two loops $P^{\,(0)\rm s}_{\,\rm ns}$ and $P^{\,(1)\rm s}_{\,\rm ns}$ both vanish.
$P^{\,(2)\rm s}_{\,\rm ns}$ is displayed in Fig.~\ref{fig:3} on the right. 
Quiet unexpectedly, $P^{\,(2)\rm s}_{\,\rm ns}$ also behaves like $\ln^{\, 4}x$ for
$x \ra 0$, and here the leading small-$x$ terms do indeed provide a
reasonable approximation. In fact, this function dominates the
small-$x$ behaviour of the non-singlet splitting functions, for
$n_{\!f}=4$ being, for example, about 7 times larger than
$P^{\,(2)\pm}_{\,\rm ns}(x)$ at $x = 10^{-4}$. 

Let us next illustrate the 
numerical effect of the three-loop splitting functions $P^{(2)}_{\rm ab}(x)$ on the evolution 
of the singlet-quark and gluon distributions 
$q_{\rm s}(x,\mu_f^{\,2})$ and $g(x,\mu_f^{\,2})$.  For all figures 
we choose a reference scale $\mu_f^{\,2} \: =\: \mu_0^{\,2}\:\simeq\:
30$ GeV$^2$ --  a scale relevant, for example, for deep-inelastic 
scattering both at fixed-target experiments and the {\it ep} collider 
HERA -- and employ the sufficiently realistic model distributions
\bea
\label{eq:shapesq}
{\lefteqn{
  xq_{\rm s}(x,\mu_{0}^{\,2}) = }}\\&&
  0.6\: x^{\, -0.3} (1-x)^{3.5}\, (1 + 5.0\: x^{\, 0.8\,})\, ,  \nn \\
\label{eq:shapesg}
{\lefteqn{
  xg (x,\mu_{0}^{\,2}) = }}\\&&
  1.6\: x^{\, -0.3} (1-x)^{4.5}\, (1 - 0.6\: x^{\, 0.3\,})\, , \nn
\eea
irrespective of the order of the expansion to facilitate the comparison 
of the LO, NLO and NNLO contributions to the splitting functions.

\begin{figure}[htb]
\begin{center}
\includegraphics[width=7.5cm]{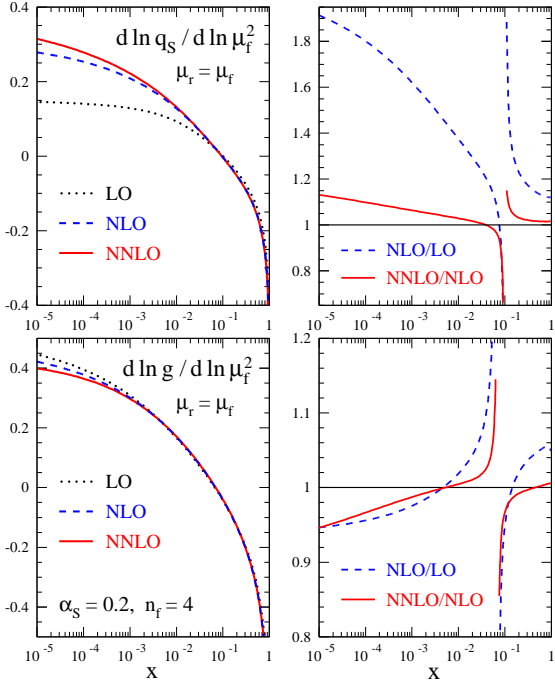}
\end{center}
\vspace*{-9mm}
\caption{The perturbative expansion of the scale derivatives~(\ref{eq:evols}) of the singlet 
distributions~(\ref{eq:shapesq}),(\ref{eq:shapesg}).}
\label{fig:4}
\end{figure}
\begin{figure}[htb]
\begin{center}
\includegraphics[width=7.5cm]{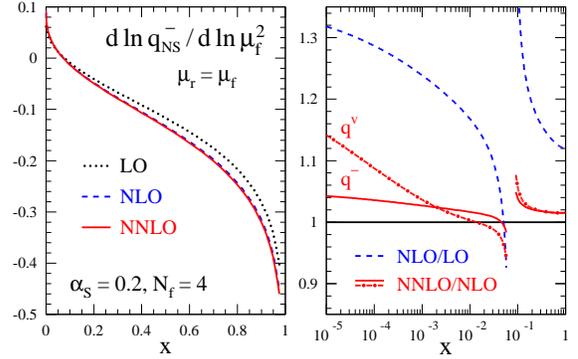}
\end{center}
\vspace*{-9mm}
\caption{The perturbative expansion of the scale derivative~(\ref{eq:evolns}) of the
non-singlet distributions $q_{\,-}$ and $q_{\,\rm v}$ for the input~(\ref{eq:shapesns}).}
\label{fig:5}
\end{figure}

In Fig.~\ref{fig:4} we show in the top row 
the perturbative expansion of the scale derivative $\,\dot{q}_
 {\,\rm s} \,\equiv\, d \ln q_{\rm s} / d\ln \mu_f^{\,2}\,$ of the
 singlet quark distribution, i.e. the top row of Eq.~(\ref{eq:evols}), 
at $\,\mu_f^{\,2} \: =\: \mu_0^{\,2\,}$ for four flavours, $\as = 0.2$, and the 
initial conditions specified in Eqs.~(\ref{eq:shapesq}), (\ref{eq:shapesg}).
The bottom row of Fig.~\ref{fig:4} shows the same for the gluon distribution $g$,
i.e. for the bottom row of Eq.~(\ref{eq:evols}).
The spikes close to $x = 0.1$ in the right parts of both figures
are due to zeros of the LO and NLO predictions and do not represent
large corrections.

The NNLO corrections are small at large $x$ with respect to both the 
total derivative and the NLO contributions.  At small-$x$ the NLO 
contributions are very large for the quark evolution. 
Consequently the total NNLO corrections, while reaching 10\% at $x = 10^{\,-4}$, 
remain smaller than the NLO results by a factor of eight or more over the full $x$-range.

In Fig.~\ref{fig:5} we show the perturbative expansion of the logarithmic scale derivative 
$\,d \ln q_{\,\rm ns}^{\, -}/ d\ln \mu_f^{\,2}\,$ for a characteristic
non-singlet quark distribution 
\bea
\label{eq:shapesns}
xq_{\rm ns}^{\, i}\, =\, x^{\, 0.5} (1-x)^3\, ,
\,\,\,\,\,\,\,\,\,\,\,\,\,\,\,\,\, i=\pm,\mbox{v},
\eea
at the standard scale $\mu_r = \mu_f$. In addition, on the right hand side of  Fig.~\ref{fig:5}, 
the non-singlet quark distribution $q_{\,\rm ns}^{\, {\rm v}}$ is also displayed.

\begin{figure}[htb]
\begin{center}
\includegraphics[width=7.5cm]{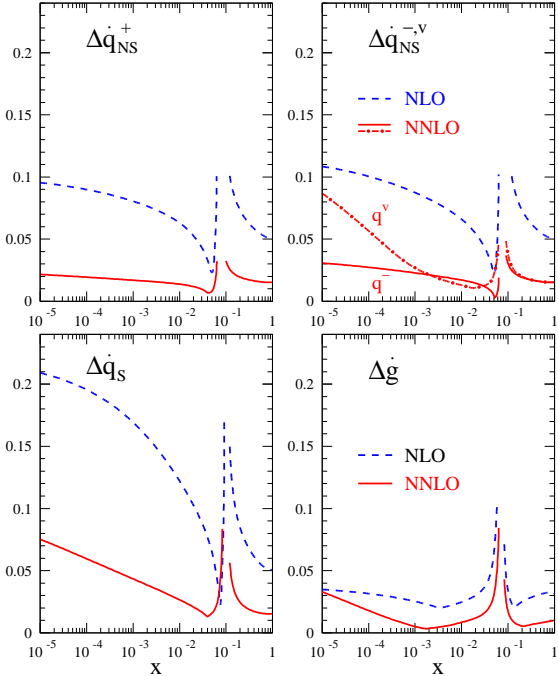}
\end{center}
\vspace*{-9mm}
\caption{The relative scale uncertainty $\Delta \dot{f}$~(\ref{eq:screl}) for all  
non-singlet and singlet cases.}
\label{fig:6}
\end{figure}

Finally, we turn to the stability of the perturbative expansions 
under variations of the renormalization scale $\mu_{\,r}$. For 
$\mu_{\,r} \neq \mu_f$ the expansion of the splitting functions in 
Eq.~(\ref{eq:Pexp}) is, using the abbreviation $a_{\rm s} \equiv 
\as/(4\pi)\,$, replaced by 
\bea
{\lefteqn{
  P_{\rm ab}(\mu_f,\mu_{\,r})
   = 
    a_{\rm s}(\mu_{\,r}^{\,2}) \, P_{\rm ab}^{(0)}  }} 
\\ && \nn
+ 
    a_{\rm s}^{\,2}(\mu_{r}^{\,2}) \, \left( P_{\rm ab}^{(1)} 
    - \beta_0\, P_{\rm ab}^{(0)} \ln \frac{\mu_f^{\,2}}{\mu_{r}^{\,2}} 
    \right) \: 
\\ &&\nn
+ 
   a_{\rm s}^{\,3}(\mu_{r}^{\,2}) \, \left( P_{\rm ab}^{(2)}
    - \bigg\{ \beta_1 P_{\rm ab}^{(0)} + 2\beta_0\, P_{\rm ab}^{(1)}
   \bigg\} \ln \frac{\mu_f^{\,2}}{\mu_{r}^{\,2}} 
\right. \\ &&\nn \left.
    \quad\quad + \beta_0^2\, P_{\rm ab}^{(0)} \ln^2 
   \frac{\mu_f^{\,2}}{\mu_{r}^{\,2}} \, \right) + \ldots \nn \:\: ,
\eea
where $\beta_k$ represent the \MSb\ expansion coefficients of the 
$\beta$-function of QCD \
\cite{Caswell:1974gg,Jones:1974mm,Tarasov:1980au,Larin:1993tp}.

In Fig.~\ref{fig:6} the relative scale uncertainties of the average results is plotted,
which is conventionally estimated by
\beq
\label{eq:screl}
 \Delta \dot{f} \: \equiv \:
 \frac{\max\, [ \dot{f}(x,\mu_r^2 
)] - \min\, [\dot{f}(x,\mu_r^2
)] }
 { 2\, |\, {\rm average}\, [\dot{f}(x, \mu_r^2 
)]\, | }\, ,
\eeq
where the scale $\mu_r^2$ varies $\mu_r^2 \in [\frac{1}{4}\mu_f^2, 4\mu_f^2]$.

The top row of Fig.~\ref{fig:6} shows the renormalization scale uncertainty of the NLO and NNLO 
predictions for the scale derivative of $q_{\rm ns}^{\,i}$, $i=\pm,\rm v$,
as obtained from the quantity $\Delta \dot{q}_{\rm ns}^{\, i}$ defined 
in Eq.~(\ref{eq:screl}). 
The bottom row displays the same, but for the singlet-quark distributions 
$q_{\rm s}$ and the gluon distribution $g$.
For all quark distribution, these uncertainty estimates 
amount to 2\% or less at $x > 10^{-2}$ 
(for the gluon distribution 1\% at $x > 3\cdot 10^{-4\,}$), an
improvement by more than a factor of three with respect to the
corresponding NLO results. 

In general, for $x \gsim 10^{-3}$ the perturbative expansion for the scale 
derivatives $\dot{f}\,\equiv\,d\ln f(x,\mu_f^{\,2})/d\ln\mu_f^{\,2\,}$, 
$f = q_{\rm ns}^i,\:q_{\rm s},\: g$ appears to be very well convergent 
and suggests a residual higher-order uncertainty of about 1\% or less at $\as \lsim 
0.2$. Consequently the perturbative evolution can be safely extended to 
considerably larger values of $\as$, hence lower scales, in this range 
of $x$. 
Larger corrections have to be expected at small $x$, 
but the results of the small-$x$ resummation alone will not help here. 
Further progress at small $x$ would require at least a four-loop generalization 
of the fixed-$N$ calculations~\cite{Larin:1994vu,Larin:1997wd,Retey:2000nq} and of
the $x$-space approximations~\cite{vanNeerven:2000wp} linking them to
the small-$x$ limits. In addition, one should also keep in 
mind that at fourth order the new colour structure 
$d^{abc\,}d_{abc}/n_c$ also will appear in singlet splitting functions.

\section{CONCLUSION}

We have calculated the complete third-order contributions to the
splitting functions governing the evolution of unpolarized 
parton distribution in perturbative QCD. 

The calculation is performed in Mellin-$N$ space and follows the 
previous fixed-integer $N$ computations~\cite{Larin:1994vu,Larin:1997wd,Retey:2000nq}.
The extension to the complete analytical $N$-dependence is the crucial new feature.
It required the set-up of an elaborate reduction scheme for the corresponding loop integrals, 
an improved understanding of the mathematics of harmonic sums, difference equations and harmonic 
polylogarithms, and finally the implementation of corresponding tools, together with other new
features \cite{Vermaseren:2002rp}, in the symbolic manipulation program
{\sc Form}~\cite{Vermaseren:2000nd}.

Furthermore, by keeping terms of order $\varepsilon^0$ in dimensional regularization 
throughout the calculation, we have also obtained the third-order coefficient 
functions for the structure functions $F_2$ and $F_L$ in electromagnetic and 
for $F_3$ in charged-current deep-inelastic scattering~\cite{MVV5}. 
Additionally, the present method can be used to generalize our fixed-$N$ three-loop 
calculation of the photon structure \cite{Moch:2001im} to all $N$ and it should
also be possible to obtain the polarized NNLO splitting functions in this manner.

The results for the three-loop splitting functions have been presented in both 
Mellin-$N$ and Bjorken-$x$ space in Refs.~\cite{Moch:2002sn,Moch:2004pa,Vogt:2004mw} 
and agree with all partial results available in the literature.
We have investigated the numerical impact of the three-loop (NNLO)
contributions on the evolution of the parton distributions.
The perturbative expansion appears to be very well convergent except for very small $x$ 
and shows good stability under variation of the scales.
Thus, with the results presented, the precision of the perturbative predictions for 
parton evolution has been greatly improvement.

\end{document}